\documentclass[twocolumn]{aastex631}

\newcommand{\bz}{\ensuremath{\langle B_z \rangle}}
\newcommand{\bs}{\ensuremath{\langle \vert B \vert \rangle}}

\newcommand{\teff}{\ensuremath{T_{\rm eff}}}
\newcommand{\fo}{\ensuremath{f_\parallel}}
\newcommand{\fe}{\ensuremath{f_\perp}}
\newcommand{\cz}{\ensuremath{C_z}}

\shorttitle{Multiple channels for the onset of magnetism in isolated white dwarfs}
\shortauthors{Bagnulo \& Landstreet}

\begin{document}

\title{Multiple channels for the onset of magnetism in isolated white dwarfs}

\author[0000-0002-7156-8029]{Stefano Bagnulo}
\affiliation{Armagh Observatory \& Planetarium \\
College Hill \\
Armagh, BT61 9DG, UK}

\author[0000-0001-8218-8542]{John D. Landstreet}
\affiliation{Armagh Observatory \& Planetarium \\
College Hill \\
Armagh, BT61 9DG, UK}
\affiliation{University of Western Ontario \\
1151 Richmond St. N \\
London, N6A 3KT, Ontario, Canada}

\begin{abstract}
The presence of a strong magnetic field is a feature common to a significant fraction of degenerate stars, yet little is understood about field origin and evolution. New observational constraints from volume-limited surveys point to a more complex situation than a single mechanism valid for all stars. We show that in high-mass white dwarfs, which are probably the results of mergers, magnetic fields are extremely common and very strong, and appear immediately in the cooling phase. These fields may have been generated by a dynamo active during the merging. Lower mass white dwarfs, which are often the product of single star evolution, are rarely detectably magnetic at birth, but fields appear very slowly, and very weakly, in about a quarter of them. What we may see is an internal field produced in an earlier evolutionary stage that gradually relaxes to the surface from the interior. The frequency and strength of magnetic fields continue to increase to eventually rival those of highly-massive stars, particularly after the stars cool past the start of core crystallisation, an effect that could be responsible for a dynamo mechanism similar to the one that is active in Earth's interior.
\end{abstract}


\section{Introduction}
Almost all stars with initial masses below about $8\,M_\odot$, whether single or in close binary systems, end their nuclear evolution by collapsing to the white dwarf (WD) state. The basic structure and the cooling history of WDs have been understood for many decades \citep{Chandra31,Mestel52}. However, the vast literature on these objects indicates that many aspects of the formation, structure and cooling of such stars have turned out to be extremely complicated. One aspect of this complication is the presence of global magnetic fields at the surfaces of some, but not all, WDs. The observed fields range in strength from about 30\,kG to several hundred MG \citep{BagLan21}. They have a structure organised at a large scale, often dipole-like, and do not appear to evolve on an observable time scale, although an observed field sometimes varies periodically due to WD rotation. 

The origin of magnetic fields in WDs is not well understood, but several ideas have been put forward to explain their presence, for example that the currently visible field is a descendant of a field that was present when the star was in a previous evolutionary stage, or that the observed field is generated by a contemporary dynamo excited in the core of a rotating WD by the convection produced by sinking solids when crystallisation begins \citep{Iseetal17,Ginetal22}. Another family of field generation mechanisms suggest that a field is produced by close binary evolution, for example by a dynamo acting during a post common envelope phase \citep{Touetal08,Brietal18Isolated}, or by a dynamo that is active during the merger of a binary pair of WDs which become a single star \citep{Garetal12}. 

The only way we know to test these theories is to consider their implications for the population of WDs, and compare predictions with observations. However, observations do not automatically give a clear view of possible correlations between the presence of a magnetic field and other characteristics of the star (such as age or mass), because of the prevalent use of output from low-resolution spectroscopy and of magnitude-limited surveys which strongly skew the sample of known magnetic WDs (MWDs) towards hot stars with field strength in the range $\sim 2-100$\,MG, as will be discussed in Sect.~\ref{Sect_Bias}. A previous attempt to avoid these biases is represented by the spectropolarimetric survey of the 20\,pc volume \citep{BagLan21}, in which more than 20\,\% of WDs were found to host magnetic fields. An important characteristics of the local 20\,pc volume is that MWDs with $M \le 0.75\,M_\odot$ and older than $\sim 2$\,Gyr are quite numerous, more than one out of four WDs, while among 20 WDs younger than 0.5\,Gyr, only one is magnetic. The 20\,pc volume still includes only a relatively small sample of WDs, none of them belonging to the group of (rare) highly massive WDs. Increasing the volume completely surveyed with spectropolarimetric techniques so as to achieve a significant increase of the survey S/N would require a huge observational effort. It is possible, however, to start the systematic exploration of certain slices of the age-mass space, and obtain a limited yet still unbiased view of the occurrence of the magnetism in, for example, specific age ranges of degenerate stars. Here we report important discoveries from a spectropolarimetric survey of the WDs younger than $\sim 600$\,Myr of the local 40\,pc volume.

\section{Observational biases in the literature}\label{Sect_Bias}
The first discovery of a magnetic field in a degenerate star was made about fifty years ago via the detection of a signal of broadband circular polarisation, which was ascribed to an effect of dichroism of the continuous opacity caused by the presence of a magnetic field of at least several tens of MG \citep{Kemetal70}. A dozen more WDs were subsequently discovered to be strongly magnetic using the same technique of detection of circular polarisation of the continuum \citep{Landstreet92}, but since the 1990s, the great majority of MWDs have been discovered via the detection of the Zeeman effect in the spectral lines, either observed with a normal spectrograph sensitive to the light intensity, or via spectropolarimetry, sensitive also to the circular polarisation of the line profiles. A few hundred WDs have been checked for magnetic fields by specifically dedicated spectropolarimetric surveys \citep[e.g.][]{SchSmi95,Putney97,Kawetal07,Kawetal12,BagLan18,BagLan21}, but the bulk of discoveries of MWDs have been obtained as a by-product of large spectroscopic surveys carried out in a broader astrophysical context, such as the Sloan Digital Sky Survey (SDSS) \citep{Kepetal13}, or of  surveys of WDs not specifically aimed at field detection \citep{Lieetal03,Napetal20}. Today, we know of more than 600 MWDs \citep{Feretal15}, more than half of which were discovered from low-resolution, mostly low S/N SDSS data, which are typically sensitive to field strength between 2 and 100\,MG \citep{Kepetal13}. Even high-resolution spectroscopy cannot detect fields weaker than a few tens kG because in that low-field regime, Zeeman splitting is washed out by intrinsic pressure broadening of the spectral lines. At the higher end of field strength (100\,MG or more), MWDs may escape detection via spectroscopy because in that regime the magnetic energy becomes comparable to the atomic Coulomb energy, and the position and strength of the components of spectral lines changes to the extent that spectra become difficult to recognise and classify \citep{Wunetal85}.  In all WDs that are sufficiently cool to have featureless spectra, the magnetic fields are totally undetectable via spectroscopy (unless their atmospheres are polluted by metal elements). However, due to the dichroism of the continuous opacity \citep{Kemp70b}, starlight is polarised when a strong field is present, hence strongly magnetised WDs may be detected via spectro- or broadband polarimetry even when spectroscopically they would appear featureless or unrecognisable. 

Circular spectropolarimetry allows us to measure the longitudinal component of the magnetic field, averaged over the stellar disk, with a sensitivity that in WDs with deep lines may be as good as a few hundred Gauss (if the S/N is sufficiently high), and is the technique of choice to detect stellar magnetic fields. However, because several large systematic spectroscopic surveys are continuously run by large research groups, while spectropolarimetric surveys are much less common, and require much higher S/N than spectroscopy, our knowledge of the magnetism of WDs is very strongly skewed towards MWDs with deep spectral lines and with field strength between 1 and 100 MG.
\begin{figure*}
\begin{center}
\includegraphics[angle=0,width=10.0cm,trim={1.0cm 5.0cm 1.3cm 2.9cm},clip]{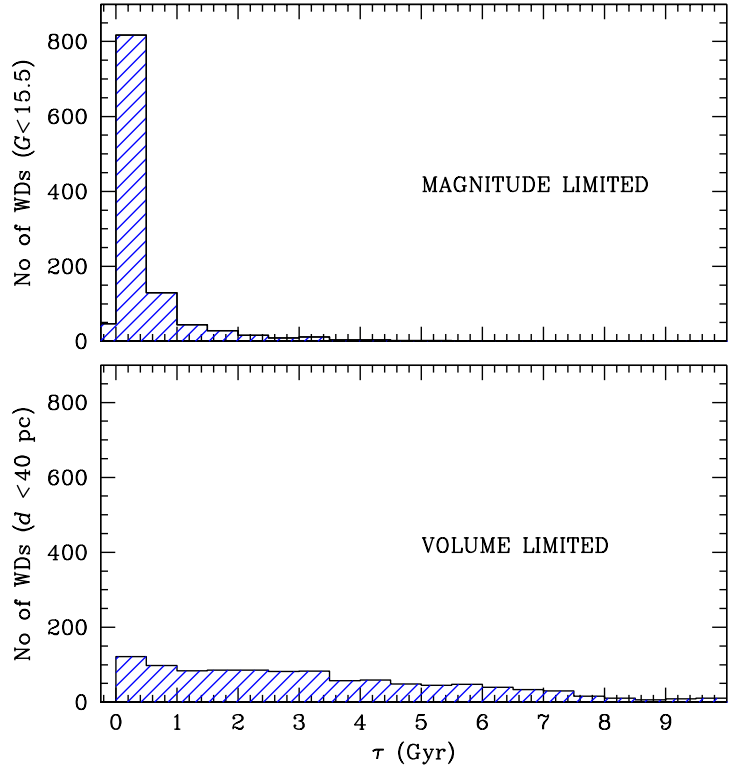}
\end{center}
\caption{\label{Fig_Limited} Age distribution of a magnitude limited sample (top panel) and a volume limited sample (bottom panel) of similar size ($\sim 1000$ WDs). Cooling age was estimated using the mass and temperature from the most recent WD catalogue of \citet{Genetal21} assuming that all stars are H-rich, and interpolating the age from the table of the Montreal group \citep{Bedetal20} using the results for thick hydrogen layer models.}
\end{figure*}

Another important bias comes from the fact that large surveys are often magnitude-limited. This kind of bias is quantified in Figure~\ref{Fig_Limited}, which shows the histograms of the age distribution of the sample of all WDs present in the local 40\,pc volume, and of the age distribution of a sample of similar size of all WDs with magnitude $G < 15.5$. It clearly appears that young WDs, much brighter than older WDs, are vastly over-represented in magnitude limited surveys. We find that 77\,\% of the WDs of the magnitude-limited sample are younger than 0.5\,Gyr, while they represent only 11\,\% of the volume-limited sample. Stars older than 3.5\,Gyr make up 40\,\% of the WDs within 40\,pc from the Sun, while they represent only 1.5\,\% of the WDs brighter than $G=15.5$. Furthermore, due to the way spectral features change with temperature (hence cooling age), magnetic fields are detected with a higher sensitivity in younger WDs than in older WDs. As a result, collecting spectroscopic measurements from magnitude-limited surveys may reveal in which kind of star field detection techniques are most effective, rather than the kinds of star in which fields are more frequently present. The ratio of known MWDs of a certain age to the total number of known MWDs may not reflect at all the way in which the frequency of the occurrence of magnetic fields varies with age. 

\section{New observations}
Gaia photometry and parallaxes allow the identification of 1077 WDs (with a probability of WD nature $p>0.75$) in the volume of space within 40\,pc from the Sun \citep{Genetal21}, and provide estimates for their temperatures and masses. We have then estimated the ages of the stars of this sample by interpolating the Montreal WD evolutionary sequence tables by \citet{Bedetal20}, and finally identified about 145 WDs younger than 600\,Myr, including about 20 WDs that are within the local 20\,pc volume and that were already discussed in a previous survey \citep{BagLan21}. A literature search showed that 11 of the young WDs within 40\,pc from the Sun had already been found to be magnetic by previous work, and that another 64 had been observed in spectropolarimetric mode and found non-magnetic (details are given in App.~\ref{Sect_Stars}).

For our new observations we have used the FOcal Reducer and low dispersion Spectrograph (FORS2) instrument of the ESO Very Large Telescope \citep{Appetal98}, the Echelle SpectroPolarimetric Device for the Observation of Stars (ESPaDOnS) instrument of the Canada-France-Hawaii telescope \citep{Donetal06}, and the Intermediate-dispersion Spectrograph and Imaging System (ISIS) on the William Herschel Telescope. With these instruments we obtained 107 new spectropolarimetric observations of 85 young WDs between 20 and 40\,pc from the Sun. Among them, 58 had never been observed in polarimetric mode before; for the majority of the remaining 27 stars, we improved the sensitivity of the measurements typically by at least a factor of 5. Only a dozen stars, mostly with ages in the range 0.5--0.6\,Gyr, were left unchecked for magnetic field. Observing strategy, data reduction, and results for individual stars are described in detail in App.~\ref{Sect_NewObs}, where we report the discovery of a strong magnetic field in the high-mass star WD\,1008$-$242 = UCAC4 328-061594,  a confirmed detection of a weak field in the high-mass WD\,0232+525 \citep{BagLan21}, a probable but to be confirmed detection of a weak field in the average mass star WD\,1704$+$481.1, and no field detection in any of the remaining 82 young WDs.

\section{Results}
Our new spectropolarimetric observations, combined with previous data collected in volume-limited samples, show that $\sim 10$\,\% of WDs younger than 0.6\,Gyr and within 40\,pc from the Sun are magnetic. Remarkably, the mass distribution of the young MWDs is highly skewed towards the highest values. Although $\sim 0.6$\,\% of the WDs of the 40\,pc volume younger than 600\,Myr have $M \ge 1.1\,M_\odot$, more than half of the young MWDs of the same volume are in that mass range, and ten out of the 12 (or 13) young MWDs have $M \ge 0.75\,M_\odot$. The average mass of the MWDs younger than 0.6\,Gyr is $0.98\,M_\odot$ (with a scattering of $\sim 0.26\,M_\odot$), while the average mass of the nMWDs in the same age range is $0.62 \pm 0.14\,M_\odot$. This is in stark contrast to what is observed in stars with cooling age $\ge 2$\,Gyr, among which the average mass of MWDs is practically indistinguishable than that of nMWDs ($0.65 \pm 0.12\,M_\odot$ and  $0.62 \pm 0.13\,M_\odot$ for the MWDs and nMWDs, respectively). Previous literature has repeatedly highlighted that the average mass of MWDs is higher than the average mass of nMWDs \citep{Liebert88,Kepetal13,McCetal20}. The new results provide a clearer view: young MWDs are generally highly massive, while magnetic fields appear in lower mass WDs only when they are older. Therefore, the average mass of younger MWDs is higher than the average mass of older MWD.

\begin{figure*}
\begin{center}
\includegraphics[angle=0,width=12.0cm,trim={1.5cm 6.5cm 0.7cm 2.9cm},clip]{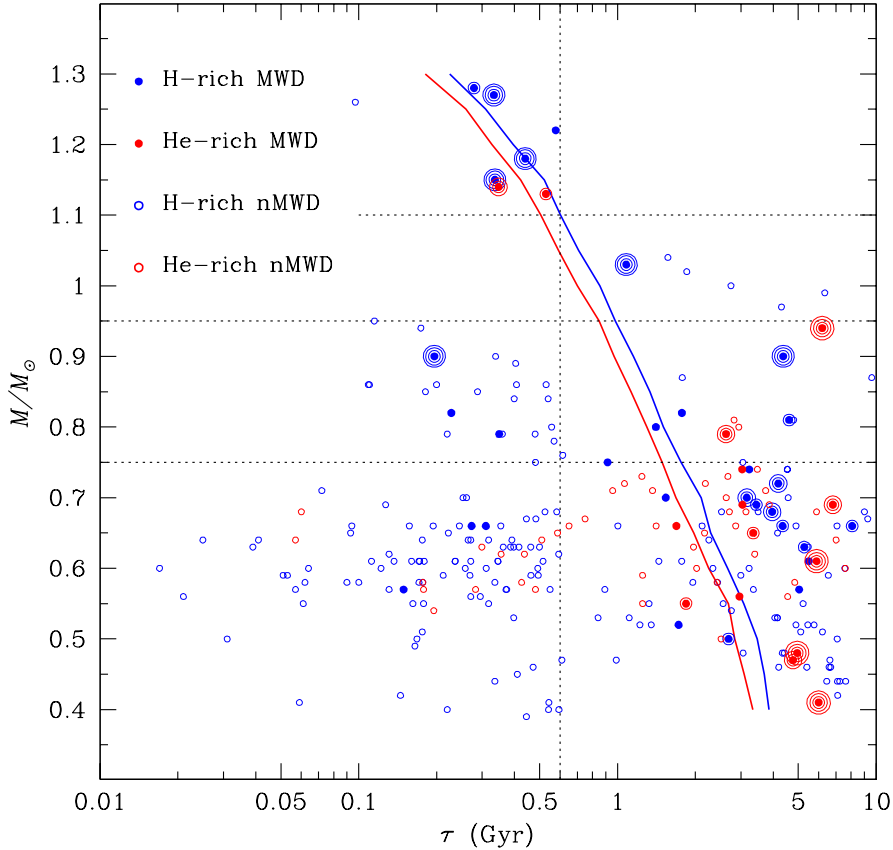}
\caption{\label{Fig_AgeMass} Age-mass diagram using data obtained for all WDs of the local 20\,pc volume and most of the WDs younger than 0.6\,Gyr up to 40\,pc from the Sun. Stellar parameters are obtained as explained in App.~\ref{Sect_Stars} and given in Table~\ref{Tab_Stars}. Filled dots identify MWDs with fields below 1\,MG; filled dots surrounded by one circle represent MWDs with field strength between 1 and 10\,MG, filled dots surrounded by two circles represent MWDs with field strength between 10 and 100\,MG, and filled dots surrounded by three circles represent MWDs with field strength $\ge 100$\,MG. Solid curves represent the onset of crystallisation; the red line refers to thin hydrogen layer models, and the blue line to thick hydrogen layers models as provided by theoretical computations \citep{Bedetal20}. The vertical dotted line marks the age limit of the 40\,pc survey; the horizontal dotted lines mark the mass boundaries discussed in the text.}
\end{center}
\end{figure*}

Figure~\ref{Fig_AgeMass} shows the position of all WDs of the 20\,pc volume plus all observed young WDs of the local 40\,pc volume (for a total of 264 stars), in a diagram of mass versus cooling age $\tau$. Notice that although different volumes are combined, each half of the diagram refers to a volume-limited sample. 

The picture that we infer is that, among the WDs younger than 0.6\,Gyr, almost all those with $M>1.1\,M_\odot$ are magnetic ($\sim 85$\,\%), and almost all those with $M \le 0.75\,M_\odot$ are non-magnetic ($\la 2.5$\,\%).  Furthermore, the field strengths of all high-mass MWDs except one are very high (ranging from $\sim 3$ to 300\,MG), while the fields of the few lower mass MWDs are much weaker in strength, mostly of order of tens of kG, and stay weak for a long time: among lower mass WDs, the youngest one that shows a MG field is almost 2\,Gyr old. In the intermediate mass regime between 0.75 and $0.95 M_\odot$, three out of 20 young WDs are magnetic. Two of these MWDs have very weak fields, one has a very strong field. These stars may belong to an overlap region between lower (normal) mass, mostly nMWDs, and high-mass, mostly magnetic WDs. The star with the strongest field (700\,MG) has $M=0.9\,M_\odot$ and may well belong to the category of high-mass, high-field strength WDs. The mass range 0.95 to 1.10\,$M_\odot$ is not probed by our observations in young stars, and clearly more data are needed to better sample the transition mass region, as current data leave a certain degree of arbitrariness in the definition of the mass threshold values.

\section{Interpretation}
The fundamental discovery emerging from the new data is the clear identification of two populations distinguished by very different typical masses, that exhibit entirely different evolutions of magnetism. Among the most massive WDs, which represent a tiny minority of all WDs, large magnetic fields are frequent and emerge to the stellar surface immediately or shortly (within hundreds of Myr) after the start of the cooling phase, with typical strengths that are among the largest observed in WDs, often above 100\,MG. Among the young WDs with masses below $\sim 0.75 M_\odot$, values that are perfectly typical of the large majority of WDs, magnetic fields are extremely rare, down to field detection limits of a few kG, and the few fields found are among the weakest ones found in degenerate stars. The strength and frequency of the surface fields of lower mass WDs grow slowly with time: among lower mass WDs, the youngest one that shows a MG field is almost 2\,Gyr old; then 11 out of 64 WDs older than  3\,Gyr  with $M \le 0.75\,M_\odot$ have a magnetic field with strength in the same range as that typical of high-mass, young strongly magnetic WDs. Old, weakly magnetic WDs are also present, and may well be more numerous than found, because weak fields are difficult to detect in cool WDs, unless strong metal lines are present.  It is remarkable how well our new, statistically strong, and evolutionarily clear picture agrees with the trends extracted from very heterogeneous data by \citet{ValFab99}, who had suggested already the existence of two populations of MWDs similar to those found in this survey.

\begin{figure*}
\begin{center}
\includegraphics[angle=0,width=12.0cm,trim={1.8cm 6.5cm 0.9cm 2.9cm},clip]{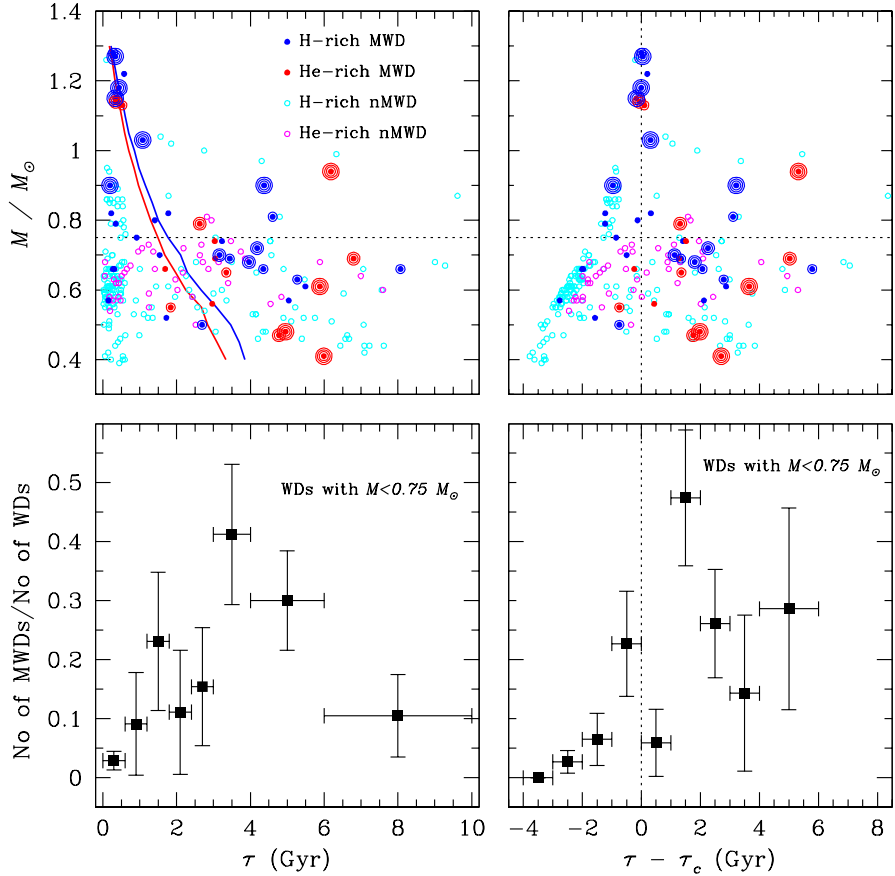}
\end{center}
\caption{\label{Fig_ctau} {\it Top panels:} Magnetic and non magnetic WDs in a age-mass diagram, with age expressed in a linear scale. The meaning of the symbols is the same as in Fig.~\ref{Fig_AgeMass}. {\it Bottom panels:} the observed distribution of the ratio between MWDs and total number of WDs with $M\le 0.75\,M_\odot$. The left panels refer to the cooling age, while in the right panels the age is counted from the onset of crystallisation, which varies from star to star according to mass.}
\end{figure*}

Mass is normally a conserved quantity as a single WD cools, so the two groups maintain their identities as they cool. The evolution of the frequency of the occurrence of MWDs in these two groups of stars with time is visualised in the left panels of Fig.~\ref{Fig_ctau}.  Now we consider how these two very different evolution paths may have arisen.

The most massive MWDs are often supposed to have been formed by an evolution pathway involving the merger of two WDs created during the evolution of a close binary system \citep{Feretal97,Dobetal12}. It has been argued that such a merger could lead to the generation of a very strong magnetic field, and it seems plausible that such a field would almost immediately be present at the WD surface \citep{Garetal12}. Furthermore, this merger event would generally be expected to leave a strong footprint in the form of a very rapid rotation of the resulting star, a feature that is found in several massive MWDs \citep{Baretal95,SchNor91,Pshetal20}. An alternative formation route for the most massive MWDs might be through single-star evolution from the most massive main sequence stars that collapse to become WDs at the end of their lives. The detection of strong magnetic fields in three massive WDs belonging to young clusters has led to the claim that intermediate-mass main sequence stars (with $M \ge 5\,M_\odot$) have a high probability of producing relatively massive MWDs \citep{Ricetal19,Caietal20}. However, this hypothesis does not find support in the larger number of spectroscopically observed WDs of $M \approx 1\,M_\odot$ in other young clusters for which no magnetic fields are reported \citep{Cumetal18}, and which are also very probably produced by single star evolution. In addition, it is unclear why such a mechanism would frequently or rapidly generate a magnetic field in WDs descended from stars with main sequence mass higher than a certain value, but would cease to be effective in WDs descended from less massive progenitors. Therefore we provisionally adopt the binary merger hypothesis for the formation of the most massive and strongly magnetic WDs of our sample. 

We next consider the possible origin of the rare weak fields in the main sample of the youngest normal-mass WDs, in which magnetism is so rare and so weak. In such stars, there does not seem to be any proposed dynamo mechanism acting in the interior of the WDs, at least until they cool enough for core crystallisation to begin. We deduce that the fields that begin to appear very weakly in young WDs, and appear more and more strongly as time goes on during the first 2 or 3\,Gyr, are probably fields left from earlier stages of evolution in the interior of newly formed normal-mass WDs. If this is the case, it appears that we may be witnessing the relaxation of a pre-existing field buried in the interior of a WD during its evolution as a main sequence or giant star. The relaxation time should probably be of the order of the global field decay timescale, which is estimated of the order of 2\,Gyr \citep{Fonetal73}. This time scale would roughly agree with the time scale over which we find fields emerging to the surfaces of MWDs. The way in which such relaxation might occur has been studied using numerical MHD simulations by \citet{BraSpr04}, who have shown that an arbitrary initially unstable, complex  field relaxes rather rapidly to a mixture of an interior toroidal field and a global poloidal (dipolar) field, which is stable. The dipolar field, which is what appears at the surface, continues to increase in strength even though the global interior field is decreasing due to resistivity.

If this picture is correct, we have not solved the puzzle of the origin of the fields observed in some normal-mass WDs, but merely displaced the origin to an earlier or later mechanism. One possible field origin may be the magnetic fields apparently detected by using asteroseismology tools on Kepler photometry \citep{Steetal16} in the cores of some red giants. Another possible source could be a dynamo operating in a shell undergoing fusion \citep{KisTho15}. In contrast, the nearly total lack of fields in the youngest normal-mass WDs may pose a serious problem for one of the oldest theories of WD field origins, the retention (freezing) and amplification of the magnetic flux observed in the atmospheres of the (chemically peculiar) Ap and Bp main sequence stars \citep{Woltjer64,Landstreet67,Angetal81}. It is not at all clear why the magnetic surface flux observed on the main sequence should retreat into the stellar interior later in evolution, only to very slowly leak back out to the stellar surface long after the formation of the WD.

It has been remarked that cool WDs that exhibit metal lines in their spectra are more frequently magnetic than any other class of WDs \citep{Holletal17,Kawetal19,BagLan19b,Kawetal21}, and this has been seen as a hint that the presence of a magnetic field is an effect of accretion from a debris disk \citep{Faretal11,KawVen14}. However, this correlation can be simply interpreted as the combination of the increase of the frequency of magnetic fields with cooling age and the fact that the presence of metal lines in cool stars enormously increases the sensitivity of the magnetic field measurements \citep{Holletal17,BagLan20}. The latter interpretation seems supported by the fact that among older WDs of the 20\,pc volume, the frequency of the occurrence of the magnetic field is similar in WDs that exhibit metal lines in their spectra, and in WDs that do not show metal lines \citep{BagLan21}. 

In recent years there has been much discussion about whether the magnetic field of WDs could be generated by a dynamo mechanism similar to the one that produces the fields of the Earth and of M dwarfs. This dynamo would be powered by the core convection that occurs when the WD degenerate core begins to crystallise in the presence of stellar rotation. Initial estimates suggested that this mechanism would require quite rapid rotation, and would at most be able to produce fields of order 1\,MG \citep{Iseetal17}. Since the original formulation of the theory, two significant modifications have been proposed that extend considerably its potential importance. First, that the field generated by this dynamo is orders of magnitude stronger than that originally estimated, in such a way to justify the observed range of field strengths \citep{Schetal21}. Secondly, that the convection is much slower than originally estimated \citep{Ginetal22}. These modifications lead to a dynamo that requires much less rapid rotation and one that readily produces fields of the order of 100\,MG, suggesting more strongly that crystallisation might play an important part in producing the fields in about 20\% of normal-mass WDs older than about $\sim 2-3$\,Gyr. 

The boundary of physical conditions under which a crystallisation-driven dynamo could begin to operate is shown in Fig.\,\ref{Fig_AgeMass} and in the top panels of Fig.~\ref{Fig_ctau} with two close oblique lines which refer, respectively, to WDs with almost no H envelope, and WDs with thick H envelopes. These two lines provide an estimate of the uncertainty of the age at which crystallisation starts for any particular WD. It is clear that some fields arise well before crossing the crystallisation line. Therefore this dynamo cannot be the only source of magnetic fields in lower mass WDs. However, the frequency of MWDs appears to increase after the crystallisation line. This is explored more quantitatively in the right-hand panels of Fig.~\ref{Fig_ctau} in which the abscissa coordinate is cooling time measured before and after the onset of crystallisation for each mass. This figure shows a steady rise in the magnetic WD fraction as a function of this time, both before and after the onset of crystallisation. The MWD frequency among normal-mass WDs does not change as the crystallisation boundary is crossed, but approximately doubles as another Gyr elapses, before falling off, perhaps because of ohmic decay, or perhaps simply because in older stars, which are generally faint and featureless, only fields stronger than a few MG may be detected. It appears that the crystallisation dynamo may well contribute significantly to the normal-mass, old, cool MWD cohort. To better understand its role, future observations should target WDs just before and after the beginning of the crystallisation phase.

\vspace{5mm}

\begin{small}
\noindent
Based on observations obtained at with the FORS2 instrument at the ESO Telescopes at the La Silla Paranal Observatory under program ID 109.235.Q001, 108.2206.001, 108.2206.002, and 0101.D-0103(C); with ESPaDOnS on the Canada-France-Hawaii Telescope (CFHT) (operated by the National Research Council (NRC) of Canada, the Institut National des Sciences de l’Univers of the Centre National de la Recherche Scientifique (CNRS) of France, and the University of Hawaii), under programmes 15BC05, 16AC05, 16BC01, 17AC01, 18AC06, 18BC02, 21BC002 and 22AC023; and with the ISIS instrument at the William Herschel Telescope (operated on the island of La Palma by the Isaac Newton Group), under programmes P15 in 18B, P10 in 19A and P8 in 19B. The observations at the CFHT were performed with care and respect from the summit of Maunakea which is a significant cultural and historic site. All raw data and calibrations of FORS2, ISIS and ESPaDOnS data are available at the observatory archives: ESO archive at {\tt https://archive.eso.org}; Astronomical Data Centre at {\tt http://casu.ast.cam.ac.uk/casuadc/}; and the Canadian Astronomical Data Centre at {\tt https://www.cadc-ccda.hia-iha.nrc-cnrc.gc.ca/en/}. This work has made use of data from the European Space Agency (ESA) mission
{\it Gaia} (\url{https://www.cosmos.esa.int/gaia}), processed by the {\it Gaia}
Data Processing and Analysis Consortium (DPAC,
\url{https://www.cosmos.esa.int/web/gaia/dpac/consortium}). Funding for the DPAC
has been provided by national institutions, in particular the institutions
participating in the {\it Gaia} Multilateral Agreement.
JDL acknowledges the financial support of the Natural Sciences and Engineering Research Council of Canada (NSERC), funding reference number 6377-2016. 

\end{small}

\bibliography{sbabib}{}
\bibliographystyle{aasjournal}

\newpage

\appendix

\section{New observations}\label{Sect_NewObs}
\begin{figure*}
\begin{center}
\includegraphics[angle=0,width=11.0cm,trim={1.9cm 7.0cm 1.3cm 2.9cm},clip]{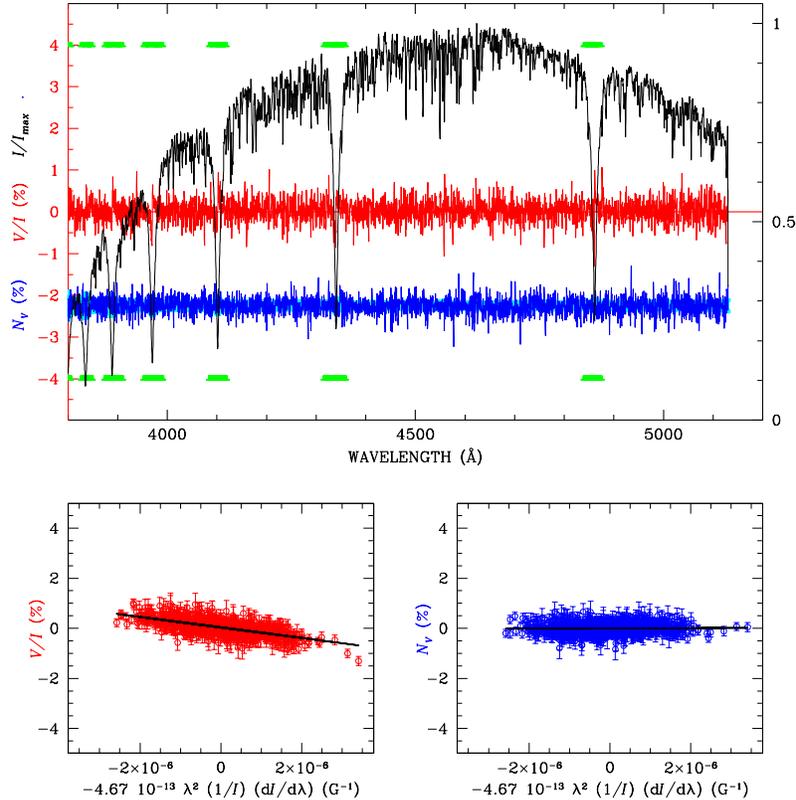}
\end{center}
\caption{\label{Fig_HD} In the upper panel, the black solid line shows the intensity profile of the magnetic Ap star HD\,94660, normalised to its maximum (its scale is given in the right axis). The shape of the intensity is due to the star's spectral energy distribution convolved with the transmission function of the atmosphere + telescope optics + instrument. The red solid line shows the $V/I$ profile (in \% units, the scale is on the left-hand axis) and the blue solid line is the null profile offset by $-2.25$\,\% for display purpose. The null profile is the difference between the $V/I$ profiles measured from two different pairs of exposures, and represents an experimental estimate of the noise. Photon-noise error bars are also shown centred around $-2.25$\,\% and appear as a light blue background. Spectral regions highlighted by green bars have been used to determine the \bz\ value from H Balmer lines. The two bottom panels show the best-fit obtained by minimising the expression of Eq.~(\ref{Eq_ChiSquare}) using the $V/I$ profiles (left panel) and the null profiles (right panel). Data were obtained with FORS2 using grism 1200B.}
\end{figure*}

\begin{figure*}
\begin{center}
\includegraphics[angle=0,width=11.0cm,trim={1.9cm 7.0cm 1.3cm 2.9cm},clip]{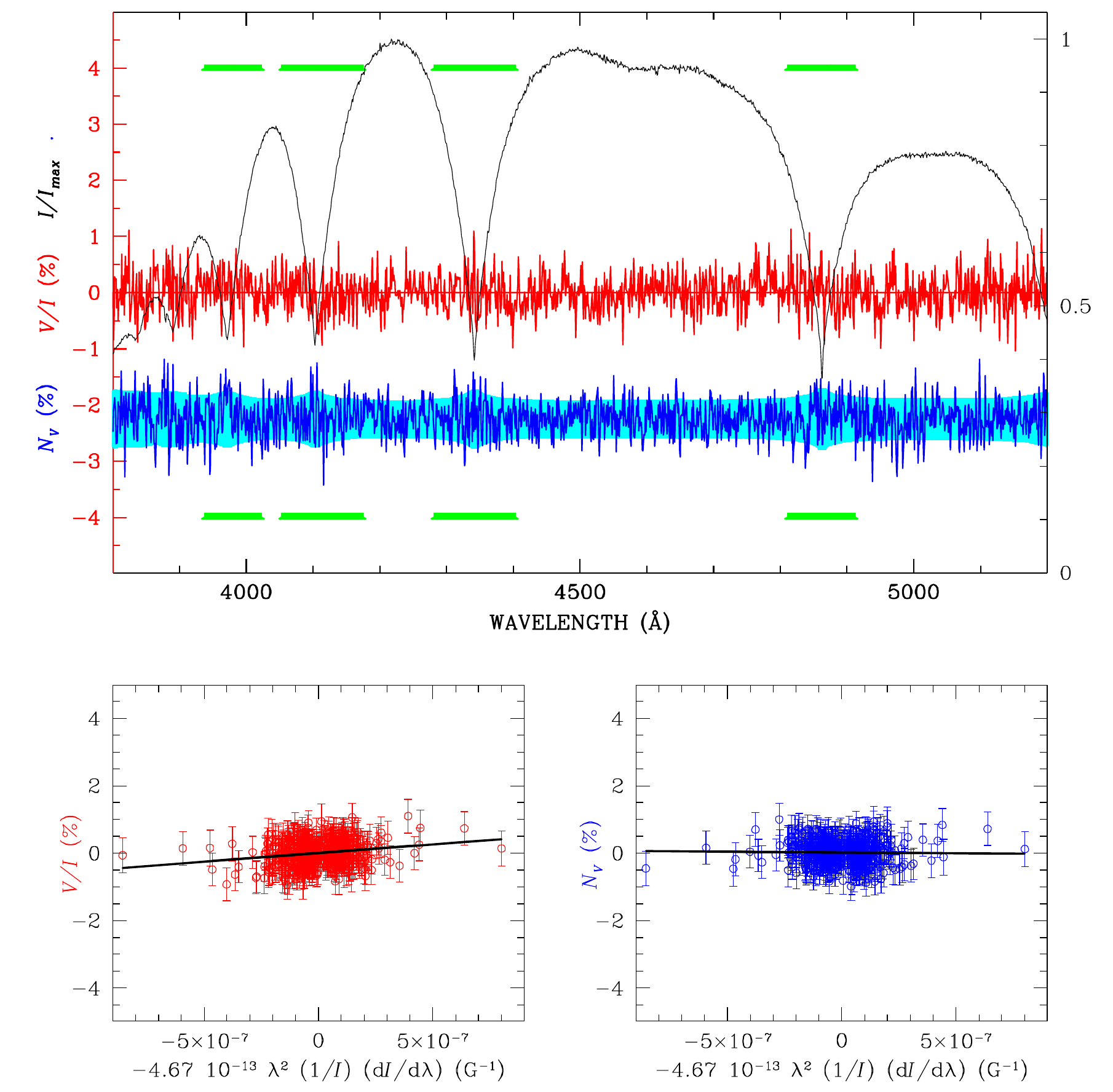}
\end{center}
\caption{\label{Fig_Example} Same as for Fig.~\ref{Fig_HD}, for the weakly magnetic star WD\,0232$+$525 = EGGR 314, observed with the blue arm of ISIS. Note the shorter range of the $x$-axis, with respect to Fig.~\ref{Fig_HD}, due to the fact that the Balmer lines of the WD are much broader than those of the main sequence Ap star HD\,94660.}
\end{figure*}

All our new observations target WDs within 40\,pc from the Sun and younger than 0.6\,Gyr, identified as explained in App.~\ref{Sect_Stars}. Observing strategy and data reduction are described in detail in numerous previous papers \citep{BagLan18,Lanetal15,Lanetal17}, and so are the methods used to measure the longitudinal field from polarised spectral lines \citep{Lanetal17,BagLan18,LanBag19a}. Briefly, the beam-swapping technique is used to minimise the impact of instrumental polarisation \citep{Bagetal09}, typically taking four exposures with the retarder waveplate at position angles $\alpha=-45^\circ$, $+45^\circ$, $+45^\circ$, $-45^\circ$. For data obtained with ISIS and FORS2, bias-subtraction, background subtraction, flux extraction and wavelength calibration were performed using standard IRAF routines, while the reduced Stokes $V/I$ profiles were obtained with simple Fortran routines by combining the various beams according to 
\begin{equation}
\begin{array}{rcl}
 \frac{V}{I} &=& \frac{1}{2} 
\Bigg\{ \left(\frac{\fo - \fe}{\fo + \fe}\right)_{\alpha=-45^\circ} -
        \left(\frac{\fo - \fe}{\fo + \fe}\right)_{\alpha=+45^\circ}
\Bigg\} \ ,
\label{Eq_Diff}
\end{array}
\end{equation}
where \fo\ and \fe\ are the flux measured in the parallel and perpendicular beam of the beam splitting device (a Wollaston prism in FORS2 and a Savart plate for ISIS), respectively. ESPaDOnS data were reduced by the pipeline LibreEsprit \citep{Moretal08}. The uncertainty of the $V/I$ profile in a spectral bin is approximately given by the inverse of the signal-to-noise ratio accumulated in that spectral bin adding up the fluxes measured in both beams at all positions of the retarder waveplate \citep{Bagetal09}. Field measurements were then obtained by minimising the expression:
\begin{equation}
\chi^2 = \sum_i \frac{(y_i - \bz\,x_i - b)^2}{\sigma^2_i}\; ,
\label{Eq_ChiSquare}
\end{equation}
where, for each spectral point $i$,
$y_i = V(\lambda_i)/I(\lambda_i)$,
$x_i = -g_\mathrm{eff} \cz \lambda^2_i \,(1/I_i\ \times \mathrm{d}I/\mathrm{d}\lambda)_i$,
and $b$ is a constant introduced to account for possible spurious polarisation in the continuum,
$g_\mathrm{eff}$ is the effective land\'{e} factor, and
\begin{equation}
\cz = \frac{e}{4 \pi m_\mathrm{e} c^2}
\ \ \ \ \ (\simeq 4.67 \times 10^{-13}\,\mathrm{\AA}^{-1}\ \mathrm{G}^{-1})
\end{equation}
where $e$ is the electron charge, $m_\mathrm{e}$ the electron mass, and $c$ the speed of light.
The alignment of the polarimetric optics was checked by measuring the magnetic field of well known magnetic stars, WD\,1900$+$705 in case of ISIS, and the Ap star HD\,94660 in case of FORS2, which both display a nearly constant longitudinal field. For example, we know that HD\,94660 shows a longitudinal field, measured from the H Balmer lines, of $\sim -2.0$\,kG \citep{Bagetal02}. We observed this star on night 2021-12-28 and measured $\bz = -2.09 \pm 0.01$\,kG. The spectral analysis of HD\,94660 is illustrated in Fig.~\ref{Fig_HD}, while an example with a WD is shown in Fig.~\ref{Fig_Example}. We note that in both cases, no obvious Zeeman effect is visible in the intensity profiles of the spectral lines, but the presence of a magnetic field is revealed by the analysis of their circular polarisation.

\begin{figure*}
\begin{center}
\includegraphics[angle=0,width=11.0cm,trim={0.9cm 5.5cm 0.3cm 3.0cm},clip]{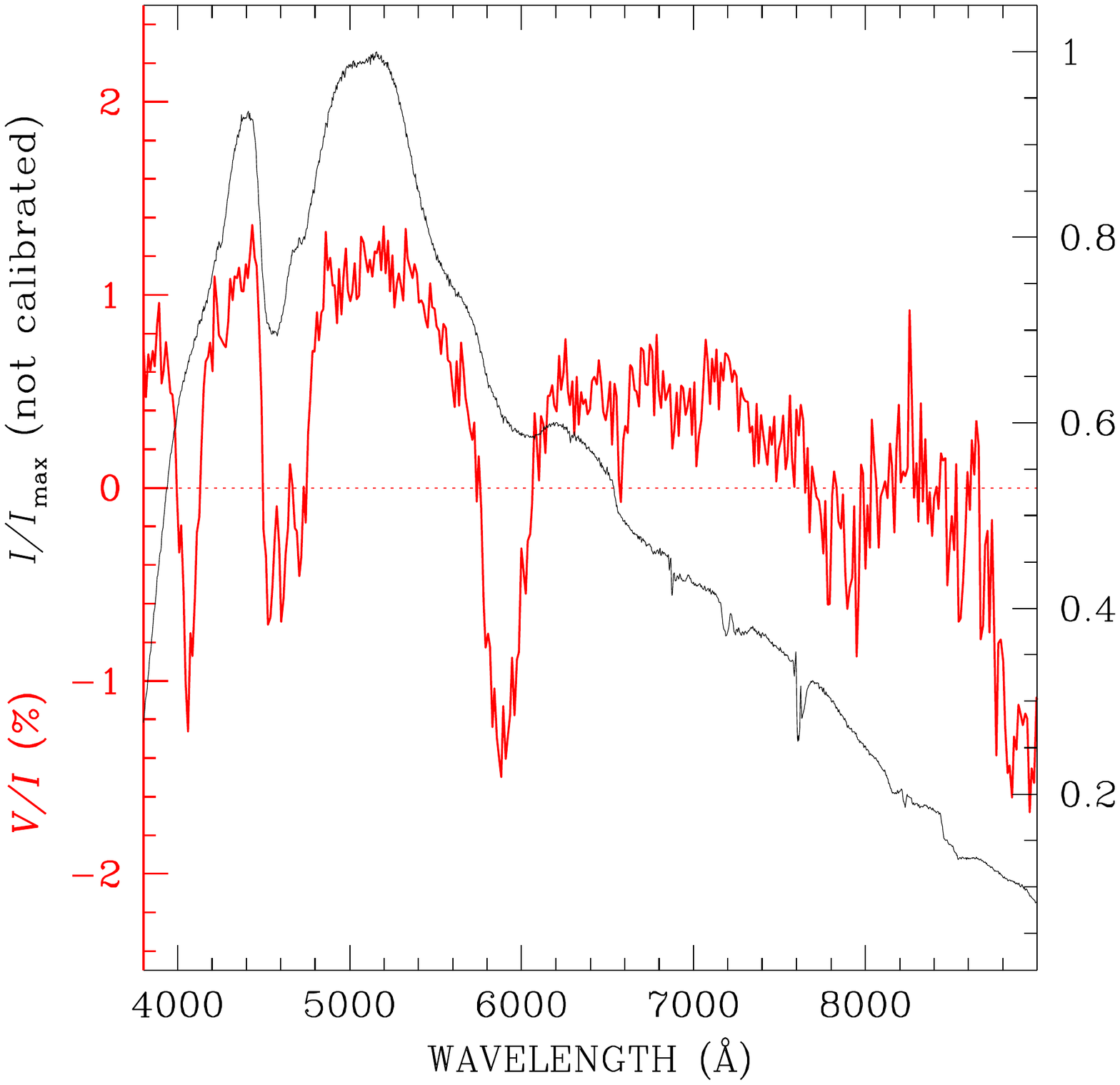}
\end{center}
\caption{\label{Fig_WD1008} FORS2 observations of WD\,1008$-$242 obtained with grism 300V. The black solid line shows the non-calibrated flux, normalised to its maximum value (scale on the right $y$-axis) and the red solid line is the circular polarisation expressed in percent (scale on the left $y$-axis).}
\end{figure*}

The observing log is given in Table~1. We confirm the discovery of a magnetic field (with ISIS and ESPaDOnS) in WD\,0232$+$525, a discovery that was already anticipated in a previous work \citep{BagLan21}, and of a signal of circular polarisation in the continuum of WD\,1008$-$242 of order 1\,\% with FORS2 (see Fig.~\ref{Fig_WD1008}), that we interpret with the presence of a $\sim 150$\,MG magnetic field \citep[based on some semi-empirical rules that associate the fraction of circular polarisation to the magnetic field strength][]{BagLan20}. Detection of a weak field was obtained with ESPaDOnS on star WD\,1704$+$481.1, but this detection definitely needs to be checked again with future observations. No field was found in any of the remaining observed stars (see Table~\ref{Tab_Log}).

\section{Data used in this work}\label{Sect_Stars}
All data used in this paper are shown in Table~\ref{Tab_Stars} and include two volume limited samples of WDs that have been checked for the presence of a magnetic field, which means that either a field has been firmly detected either via spectroscopic or spectropolarimetric techniques, or highly-sensitive field measurements have been performed by spectropolarimetric techniques, resulting into non-detection. The first sample includes virtually all WDs within 20\,pc from the Sun, already analysed by \citet{BagLan21}. The second sample, which was newly obtained for this work, is a nearly complete extension to the local 40\,pc volume of the population of WDs younger than 0.6\,Gyr, the list of which was identified with the help of the catalogue by \citet{Genetal21}. All together, Table~\ref{Tab_Stars} includes $\simeq 99$\,\% all of WDs of the local 20\,pc volume and $\simeq 90$\,\% of WDs younger than 0.6\,Gyr within 40\,pc from the Sun. 

All stellar ages (col.~8) were estimated by us using the online cooling tables of the Montreal group \citep{Bedetal20}, and a two-dimensional logarithmic interpolation on the effective temperature (given in col.~6) and mass (given in col.~7) deduced from Gaia photometry by \citet{Genetal21}. We used thick hydrogen layer models for H-rich atmospheres, thin hydrogen layer models for He-rich atmospheres -- the chemical composition of the atmosphere (col.~5) was obtained from the literature or from visual inspection of our spectra. Column~10 shows the estimate of the star's average mean field modulus \bs\ obtained as explained by \citet{BagLan21}; a zero value means that field was not detected even with polarimetric techniques. Literature (identified thanks to the SIMBAD database) and observatory archives were searched for field detection to complement our new dataset of observations of WDs -- references to field measurements are given in col.~11. The Montreal White Dwarf Database \citep{Dufetal17} was also used for the analysis of various individual stars. We note that in the previous 20\,pc volume survey by \citet{BagLan21}, stellar parameters were estimated using the various modelling results available in the literature, many of which pre-dated Gaia. These data have been revised as explained above, therefore some parameters reported in Table~\ref{Tab_Stars} may slightly differ from those given in Table~2 of \citet{BagLan21}.

In our compilation we found it useful to stick to a simple and homogeneous nomenclature of the stars, and in col.~1 of both Table~\ref{Tab_Log} and \ref{Tab_Stars} we use the naming system based on 1950 coordinates created at the Villanova University, which was originally introduced by \citet{McCSio77}, and more explicitly defined by \citet{McCSio99} as "WD" followed by the first four digits of the right ascension, the sign of the declination, the first two digits of the declination, and a third digit in which minutes of declination are expressed as the truncated fraction of degree. Some names were given incorrectly in the past, frequently because the third digit of declination was rounded instead of being truncated, but we have not changed that name if it had been already used in previous literature. Many stars of Table~\ref{Tab_Stars} have been baptised in the Villanova system for the first time in this work, and the Simbad main identifier can be found in col.~2 of Table~\ref{Tab_Stars}. Magnetic WDs are identified with the symbol "H" in their spectral classification of col.~3.

\clearpage

\begin{center}
\tabcolsep=0.14cm
\begin{longtable}{llrcccrr@{$\pm$}l}
\caption{\label{Tab_Log} New magnetic field measurements of young WDs between 20 and 40\,pc.}\\
\hline\hline
STAR                             &            INSTRUMENT                       &            Grism/                           &            DATE                             &            UT                               &            Exp                              &            S/N                              &            \multicolumn{2}{c}{\bz}         \\            &            &            Grating                          &            yyyy-mm-dd                       &            hh:mm                            &            (s)                              &            \AA$^{-1}$                        &            \multicolumn{2}{c}{(kG)}        \\            \hline
\endfirsthead
  \caption{continued.}\\
  \hline\hline
STAR                             &            INSTRUMENT                    &            Grism/                           &            DATE                             &            UT                               &            Exp                              &            S/N                              &            \multicolumn{2}{c}{\bz}         \\            &            &            Grating                          &            yyyy-mm-dd                       &            hh:mm                            &            (s)                              &            \AA$^{-1}$                        &           \multicolumn{2}{c}{(kG)}        \\            \hline
\endhead
\hline
\endfoot

WD\,0000$-$170    &  FORS2      &  600B & 2021-08-21 & 04:45 & 3000 & 485 &$-2.58 $& 1.20 \\ 
WD\,0002$+$729    &  ISIS       & R600B & 2019-10-12 & 00:14 & 3600 & 215 &$ 2.29 $& 1.12 \\
                  &             & R1200R&            &       &      & 165 &$-0.58 $& 1.05 \\ 
WD\,0125$+$135    &  FORS2      & 1200R & 2021-08-22 & 05:08 & 2800 & 400 &$-0.64 $& 0.71 \\ 
                  &  ESPaDOnS   &       & 2021-11-29 & 05:03 & 2800 &  34 &$-4.40 $& 3.84 \\ 
WD\,0126$+$508    &  ESPaDOnS   &       & 2021-08-29 & 13:16 & 3000 & 145 &$ 0.06 $& 1.10 \\ 
WD\,0133$-$116    &  FORS2      & 1200R & 2021-08-23 & 04:37 & 2800 & 440 &$ 0.25 $& 0.70 \\ 
WD\,0134$+$833    &  ESPaDOnS   &       & 2016-10-14 & 08:39 & 3360 & 170 &$-0.82 $& 0.57 \\
\hline
\end{longtable}
\end{center}

\newpage
\begin{tiny}
\tabcolsep=0.06cm
\begin{longtable}{lllrcrllrrl}
\caption{\label{Tab_Stars} Physical parameters of the WDs checked for magnetic 
field in the context of a volume-limited survey. }\\
\hline\hline
\multicolumn{11}{c}{}\\[-2mm]
\multicolumn{2}{c}{}                   & Spectral      &       &    &  \teff  & $M$      & Age   &  $d$   &$\overline{\bs}$&               \\
\multicolumn{2}{c}{STAR}               & Class         & $G$   &ATM &   (K)   &($M_\odot$)&(Gyr)  &  (pc)  &   (MG)   &References                     \\
\hline
\endfirsthead \caption{continued.}\\
  \hline\hline
\multicolumn{11}{c}{}\\[-2mm]
\multicolumn{2}{c}{}                   & Spectral      &       &    &  \teff  & $M$      & Age   &  $d$   &$\overline{\bs}$&               \\
\multicolumn{2}{c}{STAR}               & Class         & $G$   &ATM &   (K)   &($M_\odot$)&(Gyr)  &  (pc)  &   (MG)   &References                     \\
\hline \endhead
\hline \endfoot
\multicolumn{11}{c}{} \\ [-2mm]
{\bf WD\,2359$-$434  }& LAWD 96                  & DAH          & 12.9  & H  &   8428  & 0.82& 1.774&  8.3   &   0.10               & B\&L21              \\
{\rm WD\,0000$-$345  }& LAWD 1                   & DC           & 14.9  & He &   6332  & 0.66& 2.86 & 14.8   &   0$\phantom{.00}$   & B\&L21               \\ 
{\rm WD\,0000$-$170  }& EGGR 508                 & DB           & 14.6  & He &  12868  & 0.62& 0.36 & 38.1   &   0$\phantom{.00}$   & Ftw                  \\ 
{\rm WD\,0002$+$729  }& GD 408                   & DBZ          & 14.3  & He &  13788  & 0.63& 0.30 & 34.5   &   0$\phantom{.00}$   & Itw                  \\ 
{\bf WD\,0004$+$122  }& LP 464-57                & DCH          & 16.3  & He &   5098  & 0.69& 6.80 & 17.5   & 100$\phantom{.00}$   & B\&L21               \\ 
{\bf WD\,0009$+$501  }& EGGR 381                 & DAH          & 14.2  & H  &   6483  & 0.74& 3.23 & 10.9   &   0.25               & B\&L21               \\ 
{\bf WD\,0011$-$721  }& L 50-73                  & DAH          & 15.0  & H  &   6275  & 0.52& 1.72 & 18.8   &   0.37               & B\&L21               \\
\hline  
\end{longtable}
\end{tiny}
    \begingroup
    \let\clearpage\relax

\noindent
\begin{small}
  Key to the abbreviations used in col.~11:
   B\&L21: original references are given in Table~1 of \citet{BagLan21}.\
 Etw: this work, using the ESPaDOnS instrument; 
 Itw: this work, using the ISIS instrument;
 Ftw: this work, using the FORS instrument.\
 Azn+04: \cite{Aznetal04};
 B\&L18: \citet{BagLan18};
 Jor+07:  \cite{Joretal07},
 Kaw+07: \citet{Kawetal07};
 Koe+98:  \cite{Koeetal98}
 Lan+12: \cite{Lanetal12};
 L\&B19: \cite{LanBag19b};  
 L\&B20: \cite{LanBag20};
 Lie+77: \cite{Lieetal77}; 
 Lie+93: \cite{Lieetal93};
 S\&S95: \citet{SchSmi95};
 Sch+92: \cite{Schetal92};
 Swe+94: \cite{Sweetal74}.

\noindent
We note that star WD\,0316$-$849 = RE~J0317$-$853 =  EUVE~J0317$-$853 = V*~CL~Oct is often confused with star WD\,0325$-$857 = EQ~J0317-855 = LB~9802, a hot and young non magnetic WD with which it forms a physical VB system. 
\end{small}
\endgroup

\newpage

\end{document}